\documentclass[%
 reprint,
 amsmath,amssymb,
 aps,
 onecolumn,
]{revtex4-2}

\usepackage{graphicx}
\usepackage{dcolumn}
\usepackage{bm}
\usepackage{comment}
\usepackage{siunitx}
\usepackage{hyperref}
\usepackage[mathlines]{lineno}

\begin{document}

\preprint{APS/123-QED}

\title{Inverse problems in elastohydrodynamics}
\author{Anneline H. Christensen}
\author{Kaare H. Jensen}
\email{khjensen@fysik.dtu.dk}
\affiliation{Department of Physics, Technical University of Denmark, DK-2800 Kgs. Lyngby, Denmark}

\date{\today}

\begin{abstract}
Exploring fluid-structure interactions is essential for understanding the physical principle underlying flow control in biological and man-made systems. Traditionally, we assume that the geometry is known, and from it, the solution to the coupled elastohydrodynamic problem is determined. Solving the inverse problem -- finding the geometry that leads to a desired flow -- has received comparatively less attention. Here, we present a strategy for solving inverse hydroelastic problems. Specifically, we compute the shape of a soft channel that yields a desired flow-rate versus pressure-drop relationship. The analysis is based on low-Reynolds-number hydrodynamics and linear elasticity. We demonstrate its usefulness in understanding intercellular transport in plants and the design of check valves. The sensitivity of the algorithm to fabrication errors and other limitations are discussed.
\end{abstract}
\maketitle

\section{Introduction}
\begin{figure}
    \centering
    \includegraphics[width=17cm]{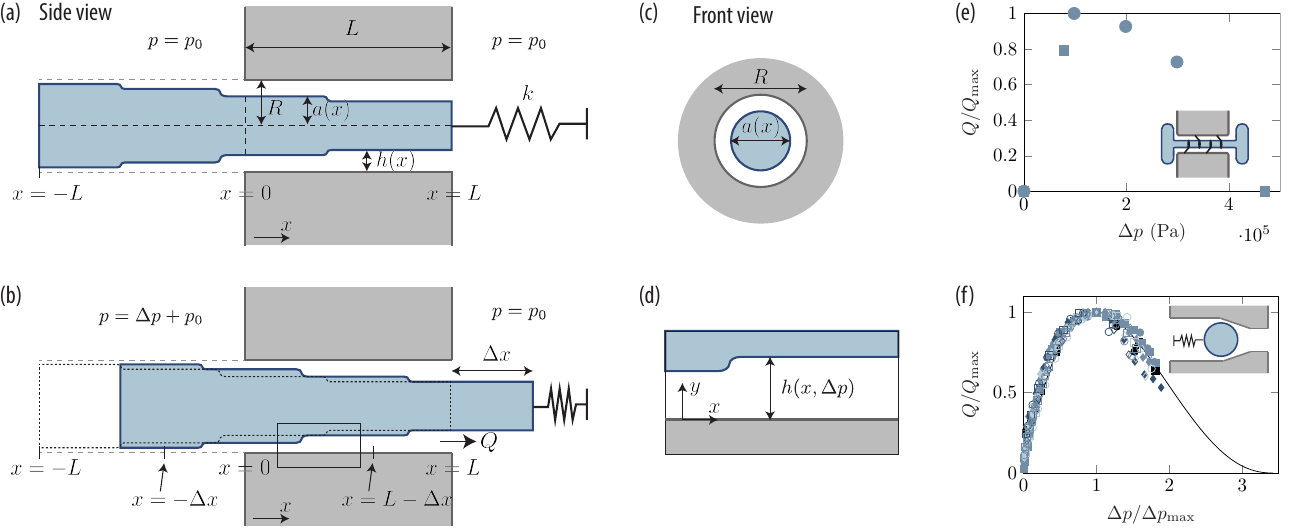}
    \caption{Diagram of the valve geometry. (a) A cylinder of variable radius $a(x)$ is connected to a spring, with spring constant $k$, inside a cylindrical pore of radius $R$ and length $L$. (b) A pressure difference $\Delta p$ across the pore drives a fluid flow $Q$ in the gap between the inner cylinder and the outer pore, $h(x,\Delta p)=R-a(x,\Delta p)$. At the same time, the inner cylinder has displaced a distance $\Delta x$ that depends on the force balance between the pressure and spring forces acting on it, which alters the geometry of the gap. (c) Front view of the cylindrical valve geometry. (d) Zoom-in of the gap height $h(x,\Delta p)$ between the inner cylinder and outer pore with the local coordinate system. (e) Experimental data of the pressure-dependent permeability through plasmodesmata pores in plants (circles from \cite{oparka1992direct} and squares from \cite{ruan2001control}). Insert shows a schematic of a plasmodesma pore. (f) Experimental data of a non-linear pressure-drop flow-rate relation from a soft valve (from previous work from our group \cite{park2018viscous}). Insert shows a schematic the soft valve from \cite{park2018viscous}.}
    \label{fig:diagram}
\end{figure}

The application of fluid-structure interactions to flow control problems is an established strategy in both nature and technology. Examples include passive regulation of vascular transport in plants and animals \cite{choat2008structure,sotiropoulos2016fluid}, and modulation of chemical reactions in microfluidic devices \cite{brett2011controlling,mosadegh2010integrated}. Despite the apparent success of passive flow regulators, the design process remains challenging \cite{oh2006review}. This is due to the non-linear nature of the governing equations, the sensitivity to initial conditions, and fluid-structure interactions that couple flow and motion in time and space. 
While the forward problem of finding the fluid flow rate $Q$ as a function of the applied pressure drop $\Delta p$ for a known channel geometry has been studied for various fluid-structure interactions (see, e.g., \cite{brandenbourger2020tunable,skotheim2004soft,alvarado2017nonlinear,park2018viscous,christov2018flow,christensen2020viscous}), the inverse problem of finding the channel geometry that will give rise to a target flow rate pressure drop characteristic is not well-studied. Here, we present a procedure to solve inverse hydroelastic problems. Given a desired pressure-drop versus flow-rate characteristic, the method can, under not too restrictive conditions, find a geometry that meets the specifiations (Fig. \ref{fig:diagram}). 


To begin our discussion on the inverse solution strategy, we must first outline the basic principles of low-Reynolds number fluid-structure interactions. When a channel contains soft or elastic elements, a pressure drop that drives fluid flow through the pore can also displace the flexible elements from their initial position. This displacement can change the pore's geometry, which then alters the pressure drop characteristics of the fluid flow. This process is ubiquitous (see references above) but a particularly simple configuration is found in the plasmodesmata channels that link adjacent plant cells  \cite{oparka1992direct,ruan2001control}.
Extending through the plasmodesmata nanopores is a string of the endoplasmic reticulum, known as the desmotubule, that blocks part of the pore and is connected by tether proteins to the cell membrane lining the pore wall \cite{nicolas2017architecture}.  \citet{park2019controlling} proposed that a pressure difference across the plasmodesma pore can cause the central desmotubule to be displaced, thereby altering the gap geometry, potentially blocking the pore to molecular trafficking. This provides a potential rationale for experimental data showing a decrease in the relative permeability with increasing applied pressure (Fig. \ref{fig:diagram} (e)) \cite{oparka1992direct,ruan2001control}. The process of a pressure drop causing a displacement of a part of the system, thereby altering the geometry and creating a non-linear pressure drop flow rate characteristic, has also been shown experimentally in a fluidic device in previous work by our team (Fig. \ref{fig:diagram} (f)) \cite{park2018viscous}. Here, a pressure drop drives the fluid flow past a sphere connected to a spring in a tapering cylindrical channel. The pressure drop, in turn, displaces the sphere from its equilibrium position, changing the geometry of the pore as a function of the pressure drop. 

Inspired by the idea that varying fluid flow-rate characteristics are seen in both nature and technology, we consider the case illustrated in Fig. \ref{fig:diagram}: viscous flow in a spring-actuated sliding valve. In particular, we aim to determine the fluid flow rate $Q$ through the pore as a function of the pressure difference $\Delta p$, and the channel geometry. Firstly, we will describe the system and create a mathematical model to determine the flow rate based on the applied pressure difference, taking into account the forces present in the system and using the lubrication approximation. We will then present our findings in two parts. Firstly, we will solve the forward problem: finding the flow-rate pressure-drop relation for a known geometry. Secondly, we will present a strategy for the inverse problem: finding the geometry that will produce a desired flow-rate pressure-drop curve. This will help us understand how to design a valve or channel to achieve a specific fluid flow characteristic. Finally, we will examine the sensitivity and conditions under which our proposed model framework is appropriate.



\section{Results}

In the following, we consider a flexible valve that links two reservoirs (Fig. \ref{fig:diagram}). Flow control occurs because the valve geometry changes shape due to the application of pressure. The valve comprises two parts. The first element is a small circular hole of radius $R$ and length $L$ in the wall that separates the two reservoirs. The second component is an inner rod of radius $a(x)$. The rod can move along the $x$ axis but the motion is resisted by a linear spring (spring constant $k$). A pressure difference across the valve, $\Delta p$, drives fluid flow along the $x-$direction in the shallow gap of height $h(x)=R-a(x)$. We assume the gap between the two cylinders is much smaller than both the radius of either cylinder, $h(x)\ll R,a(x)$ and the length of the pore $h(x)\ll L$. The equilibrium position of the textured rod is found by a balance of the spring and hydrodynamic forces acting on it. The height profile of the gap between the concentric cylinders is thus a function of both the axial position $x$ and pressure drop $\Delta p$ across the pore $h=h(x,\Delta p)$ (Figs. \ref{fig:diagram} (b) and (d)). 

The flow of a Newtonian fluid in the pore is described by the Navier-Stokes and continuity equations 
\begin{subequations}
\begin{align}
    \rho(\partial_t\mathbf{u}+\mathbf{u}\cdot\nabla\mathbf{u})&=-\mathbf{\nabla}p+\eta\nabla^2\mathbf{u},
    \\ \mathbf{\nabla}\cdot\mathbf{u}&=0,
\end{align}
\end{subequations}
where $\mathbf{u}=(u_x,u_y)$ is the velocity field, $p$ the pressure, and $\rho$ and $\eta$ the density and viscosity of the fluid, respectively. In our analysis, we assume the Reynolds number is small, i.e., that the viscous forces dominate compared to the inertial ones, so we can neglect the inertial terms of the Navier-Stokes equations and use the Stokes' equation. Further, with the assumption that the aspect ratio of the channel is small $h/L\ll 1$ and that the gap height is much smaller than the radii of the cylinders, the effects of curvature can be ignored, and we can use the two-dimensional lubrication approximation to describe the hydrodynamics in the gap. 

%
\subsection{Low-Reynolds-number flow}
%
In the steady-state and low Reynolds-number limit, the velocity field $\mathbf{u}$ and pressure $p$ in the gap between the concentric cylinders can be described using the lubrication equations, 
\begin{subequations}
\begin{align}
    \dfrac{\partial p}{\partial x}&=\eta\dfrac{\partial^2u_x}{\partial y^2},
    \\ \dfrac{\partial p}{\partial y}&=0,
    \\ \dfrac{\partial u_x}{\partial x}+\dfrac{\partial u_y}{\partial y}&=0.
\end{align}
\end{subequations}
We consider a local coordinate system, with $x=0$ at the entrance of the pore, $x=L$ at the exit, and $y=0$ at the surface of the outer cylinder positive in the normal direction towards the inner cylinder, located at $y=h(x)$ (Figs. \ref{fig:diagram} (a) and (d)). 
With no-slip boundary condition at the walls, $u_x(y=0)=u_x(y=h(x,\Delta p))=0$, the velocity in the $x-$direction is
\begin{align}
    u_x(x,y)=\frac{1}{2\eta}\frac{\partial p}{\partial x}\left(y^2-yh(x,\Delta p)\right).
\end{align}
Integrating over the cross-sectional area, the fluid flow rate is found as
\begin{align}
    Q=w\int_0^{h(x,\Delta p)}u_x(y)\mathrm dy=-\frac{w}{12\eta}\frac{\partial p}{\partial x}h^3(x,\Delta p),
\end{align}
where $w$ is the width of the channel, here $w=2\pi a$ with $a$ the radius of the inner cylinder. Integrating along the length of the pore gives, assuming the flow-rate is constant in the pore, the pressure drop flow rate relation
\begin{align}
    Q=\frac{w\Delta p}{12\eta}\left(\int_0^L h^{-3}(x,\Delta p)\mathrm dx\right)^{-1}.
\end{align}
If we write the channel height as $h(x,\Delta p)=h_0f(x,\Delta p)$, where $h_0$ is a characteristic channel height and $f(x,\Delta p)$ is a dimensionless function, the flow-rate can be written as
\begin{align}
    Q=\frac{\Delta p}{R_0}\left(\int_0^1 f^{-3}(\bar x,\Delta p)\mathrm d\bar x\right)^{-1},\label{eq:Q}
\end{align}
with $\bar x=x/L$ the dimensionless axial coordinate and the characteristic resistance
\begin{align}
    R_0=\frac{12\eta L}{wh_0^3}.\label{eq:R_0}
\end{align}
For a pore with a constant inner cylinder, i.e., $h(x,\Delta p)=h_0$, the flow rate is $Q=\Delta p/R_0$. 
\subsection{Force balance}
The flow rate depends on the displacement of the inner cylinder in the $x-$direction, which is determined by the force balance between the hydrodynamic and spring forces acting on the inner cylinder in the $x-$direction
\begin{align}
    F_{\text{spring}}+F_{\text{hyd}}=0.\label{eq:force_balance}
\end{align}
We model the compression of the spring by assuming a linear spring force
\begin{align}
    F_{\text{spring}}=-k\Delta x,
\end{align}
with spring constant $k$ and displacement in the $x-$direction from its equilibrium position $\Delta x$ (Fig. \ref{fig:diagram} (b)).

The hydrodynamic force acting on the inner cylinder can, in index notation, be written as 
\begin{align}
    F_{\text{hyd,}i}=\int_{\partial \Omega}\left[-p\delta_{ij}+\eta\left(\partial_iu_j+\partial_ju_i\right)\right]n_j\mathrm dA,
\end{align}
where $\partial\Omega$ is the surface of the inner cylinder. The pressure force acting on the inner cylinder is found by assuming the pressure is constant $p=p_0+\Delta p$ for $x<0$ and constant $p=p_0$ for $x>L$ and integrating over the cross-sectional area of the inner cylinder as
\begin{align}
    F_{\text{hyd}}^{\text{(p)}}\simeq \Delta p \pi a^2,\label{eq:F_p}
\end{align}
assuming the radius of the inner cylinder is approximately constant $a$. 
To find the viscous force acting on the inner cylinder, we use that as the pore has a length much larger than its height, $h_0/L\ll 1$, $\partial/\partial x\ll \partial/\partial y$, such that the dominant term is
\begin{align}
    F_{\text{hyd}}^{(\text{visc})}\simeq 2\pi a\eta\int_0^L\left.\frac{\partial u_x}{\partial y}\right\vert_{y=h_0}\mathrm dx=\pi ah_0\Delta p.\label{eq:F_visc}
\end{align}
By comparing the viscous force (Eq. \eqref{eq:F_visc}) and the pressure force (Eq. \eqref{eq:F_p}) we see
\begin{align}
\frac{F_{\text{hyd}}^{(\text{visc})}}{F_{\text{hyd}}^{(\text{p})}}=\frac{h_0}{a},
\end{align}
and hence in our case, when $h_0/a\ll 1$, the pressure force is the dominating term in the hydrodynamic force, and we neglect the viscous forces. 

Using the force balance from Eq. \eqref{eq:force_balance} the displacement of the inner cylinder at a given applied pressure difference $\Delta p$ is
\begin{align}
    \Delta x=\dfrac{\Delta p\pi a^2}{k}.
\end{align}
Hence, we can determine how the gap height between the concentric cylinders depends on the applied pressure difference, $\Delta p$, as
\begin{align}
    h(x,\Delta p)=h(x-\Delta x)=h\left(x-L\frac{\Delta p}{\Delta p_c}\right),\label{eq:h}
\end{align}
where we have introduced the critical pressure, $\Delta p_c$, corresponding to the pressure at which the inner cylinder is displaced one pore length, i.e., the pressure difference for which $\Delta x = L$,
\begin{align}
    \Delta p_c=\frac{kL}{\pi a^2}.\label{eq:Delta_p_c}
\end{align}
Increasing pressure thus corresponds to a negative displacement of the rod relative to its original position.
\subsection{Flow-rate versus pressure-drop characteristics for a known height profile}
Having found the relationship between the height profile of the gap between the two cylinders and the applied pressure difference, we can go back and look at the flow-rate vs. pressure-drop relation from Eq. \eqref{eq:Q}.
For a given known height profile $h(x)=h_0f(x)$, the flow rate from Eq. \eqref{eq:Q} can be found, using the substitution $\tilde x=x/L-\Delta p/\Delta p_c$, as
\begin{align}
    Q=\frac{\Delta p}{R_0}\left(\int_{-\Delta p/\Delta p_c}^{1-\Delta p/\Delta p_c}f^{-3}(\tilde x)\mathrm d\tilde x\right)^{-1}. \label{eq:Q_h}
\end{align}
The above is only valid as long as the height profile is positive, $h(x)>0$. When the gap between the two cylinders becomes zero, $h(x)=0$, the valve is closed and the flow rate is zero. 

\begin{figure}
    \centering
    \includegraphics[width=17.2cm]{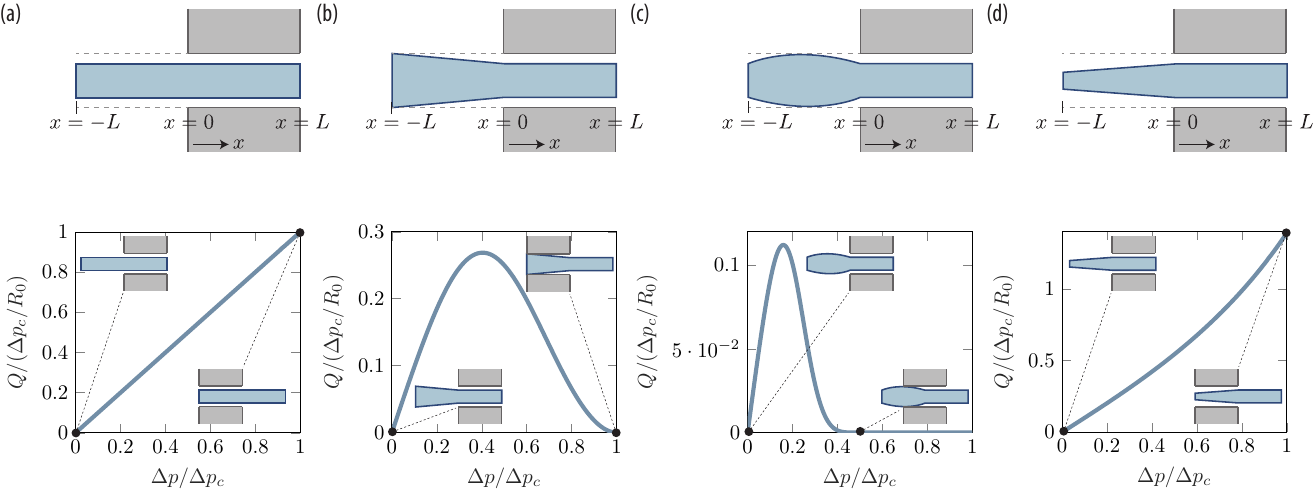}
    \caption{Pressure-drop flow-rate characteristics for known height profiles. The flow rate $Q$ (Eq. \eqref{eq:Q_h}), normalized by the flow through a pore with a constant gap height $h(x)=h_0$ at the pressure difference $\Delta p=\Delta p_c$, plotted as a function of the pressure difference $\Delta p$, normalised by the characteristic pressure drop $\Delta p_c$ (Eq. \eqref{eq:Delta_p_c}). The upper panels show schematics of the height profiles at zero pressure drop ($\Delta p=0$). The lower panels show the flow rate as a function of pressure difference, with inserts showing how the inner cylinder is displaced for various pressure drops. (a) For a constant height profile $h(x)=h_0$. (b) For a linearly decreasing height profile (Eq. \eqref{eq:h_linearly_decreasing}). (c) For a parabolic height profile (Eq. \eqref{eq:h_parabolic}). (d) For a linearly increasing height profile (Eq. \eqref{eq:h_linearly_increasing}).}
    \label{fig:fig2}
\end{figure}

To illustrate the diverse range of flow-rate characteristics possible for simple height profiles, we next consider four different examples of height profiles and their corresponding pressure-drop flow-rate characteristics (Fig. \ref{fig:fig2}). First, we look at the simplest case of a constant height profile (Fig. \ref{fig:fig2} (a)), followed by two versions of simple check valves, inspired by the experimental data from Figs. \ref{fig:diagram} (e-f) (Figs. \ref{fig:fig2} (b-c)). Lastly, we consider an example of a pore that opens instead of closes, as the check valves do (Fig. \ref{fig:fig2} (d)). 

For a channel with no variation in the height along the length of the pore, $h(x)/h_0=f(x)=1$, the flow rate increases linearly with applied pressure, and the displacement of the inner cylinder does not influence the pressure-drop flow-rate characteristics (Fig. \ref{fig:fig2} (a)). 

For the second example, we consider two simple versions of check valves, i.e., valves that close for applied pressures above a certain threshold (Figs. \ref{fig:fig2} (b-c)). 
A vegetal example relates to plasmodesmata,  small channels that link neighbouring cells in plants and allow the plant to exchange nutrients and other signaling molecules between adjacent cells. They have an approximate circular concentric geometry, with an inner structure, known as the desmotubule, blocking part of the channel \cite{nicolas2017architecture}. The desmotubule extends out of the pore and is larger than the pore diameter outside the pore. Experiments on plants have shown that the permeability of plasmodesmata changes with applied pressure (Fig. \ref{fig:diagram} (e)) \cite{oparka1992direct,ruan2001control}. Park \textit{et al.} \cite{park2019controlling} considered that an applied pressure difference across the pore can displace the inner desmotubule and potentially close the pore, and studied how the diffusive transport is influenced by the change in geometry. If the transport is governed by a pressure-driven flow, instead of diffusion, it can be modelled using Eq. \eqref{eq:Q_h}. 
We do not know the exact functional form of the gap, but show the effect of two simple height profiles leading to a closed pore: a linearly tapering height profile (Fig. \ref{fig:fig2} (b)), 
\begin{align}
    h(x) = 
    \begin{cases}
        h_0 & 0<x<L\\
        h_0\left(1+\frac{x}{L}\right) & -L<x<0
    \end{cases},\label{eq:h_linearly_decreasing}
\end{align}
and a parabolic height profile (Fig. \ref{fig:fig2} (c)),
\begin{align}
    h(x) = 
    \begin{cases}
        h_0 & 0<x<L\\
        4h_0\left(\frac{1}{2}+\frac{x}{L}\right)^2 & -L<x<0
    \end{cases}.\label{eq:h_parabolic}
\end{align}
The part of the inner cylinder that is initially inside the pore has a constant radius, i.e., the pore gap has a constant height $h(x)/h_0=f(x)=1$, for $\Delta p=0$. 

In both cases, the flow rate initially increases linearly with small applied pressure differences. As the pressure difference increases, displacing the inner cylinder, the gap height decreases, thereby decreasing the flow rate compared to the linearly increasing flow rate. When the inner cylinder is displaced, respectively, one (Fig. \ref{fig:fig2} (b)) and a half (Fig. \ref{fig:fig2} (c)) pore length, the gap height is zero, and the pore closes, not permitting any flow. 
A similar flow rate pressure drop curve is also observed experimentally in a fluidic system (Fig. \ref{fig:diagram} (f)) \cite{park2018viscous}. 

Lastly, we consider an example of a pore that opens as the applied pressure increases. If, instead of increasing linearly (as in Eq. \eqref{eq:h_linearly_decreasing}), the radius of the inner cylinder decreases linearly outside of the pore with, e.g, a height profile on the form
\begin{align}
    h(x) = 
    \begin{cases}
        h_0 & 0<x<L\\
        h_0\left(1-\frac{1}{4}\frac{x}{L}\right) & -L<x<0
    \end{cases},\label{eq:h_linearly_increasing}
\end{align}
the flow rate increases faster than the linearly increasing flow rate pressure drop profile from the constant gap height, as the gap height here is larger for increasing applied pressures (Fig. \ref{fig:fig2} (d)). 

\subsection{Flow-rate versus pressure-drop characteristics for a target flow-rate profile}
So far, we have studied pressure-drop flow-rate characteristics based on knowledge of the channel height profile. However, it is also of interest to consider the inverse problem of finding the height profile for a known target flow-rate pressure-drop curve. This problem could be of particular relevance to drug delivery, where specific flow-rates are required \cite{yang2022wearable}. 

Let the target flow rate pressure drop curve be $Q_t(\Delta p)$. We then want to compute the height profile $h(x)$, or the dimensionless version $f(\bar x)$, such that the flow rate from Eq. \eqref{eq:Q} is 
\begin{align}
    Q=\frac{\Delta p}{R_0}\left(\int_{-\Delta p/\Delta p_c}^{1-\Delta p/\Delta p_c}f^{-3}(\tilde x)\mathrm d\tilde x\right)^{-1}=Q_t(\Delta p),\label{eq:Q_Qt}
\end{align}
where we have used the substitution $\tilde x=x/L-\Delta p/\Delta p_c$. 
In the following, we assume the inner cylinder is limited to only be able be to be displaced one pore length $L$ through the pore. That is, the maximum allowed applied pressure difference is $\Delta p=\Delta p_c$. Let $f_0(\bar x)$ denote the gap height in undeformed pore, i.e., for $\Delta p=0$. Then, assuming $\Delta p<\Delta p_c$, Eq. \eqref{eq:Q_Qt} can be rewritten as
\begin{align}
    \int_{-\Delta p/\Delta p_c}^{0}f^{-3}(\tilde x)\mathrm d\tilde x+\int^{1-\Delta p/\Delta p_c}_{0}f_0^{-3}(\tilde x)\mathrm d\tilde x=\frac{1}{R_0}\frac{\Delta p}{Q_t(\Delta p)}.
\end{align}
Differentiating on both sides using Liebniz integral rule and letting $\bar x=-\Delta p/\Delta p_c$, we find the dimensionless gap height as
\begin{align}
    f^{-3}(\bar x)=f_0^{-3}(1+\bar x)+\frac{\Delta p_c}{R_0}\frac{\partial}{\partial\bar x}\left(\frac{\bar x}{Q_t(-\Delta p_c \bar x)}\right),\label{eq:f}
\end{align}
for $-1<\bar x<0$. The theory can be extended to allow for pressures displacing the inner cylinder more than one pore length by iteratively using Eq. \eqref{eq:f} with the previously found $f$ as $f_0$. 

For the (dimensionless) gap height $f(\bar x)$ from Eq. \eqref{eq:f} to be a physically realistic solution, it must be positive, continuous at $\bar x=0$ $(\Delta p=0)$, and give the target flow rate $Q_t(\Delta p)$ when inserted into Eq. \eqref{eq:Q}. These three conditions provide three additional restrictions on $f$, $f_0$ and $Q_t(\Delta p)$. First, to make sure we obtain the desired flow rate, we substitute $f$ (Eq. \eqref{eq:f}) into Eq. \eqref{eq:Q} and find
\begin{align}
    Q
    &=\frac{\Delta p}{R_0}\left(\frac{\Delta p}{R_0}\frac{1}{Q_t(\Delta p)}+\frac{\Delta p_c}{R_0}\lim\limits_{\tilde x=0}\left(\frac{\tilde x}{Q_t(-\Delta p_c\tilde x)}\right)+\int_0^1f_0^{-3}(\tilde x)\mathrm d\tilde x\right)^{-1}.
\end{align}
Hence, to get $Q(\Delta p)=Q_t(\Delta p)$ the initial channel height $f_0$ has to be normalised according to 
\begin{align}
    \int_0^1f_0^{-3}(\bar x)\mathrm d\bar x=-\frac{\Delta p_c}{R_0}\lim\limits_{\bar x=0}\left(\frac{\bar x}{Q_t(-\Delta p_c\bar x)}\right).\label{eq:f0_normalisation}
\end{align}
The above condition corresponds to the resistances at low pressures $(\Delta p =0)$ of the target flow-rate and the pore being equal. Second, the height of the gap should be continuous at $x=0$ at $\Delta p=0$, i.e., 
\begin{align}
    f^{-3}(\bar x = 0)&=f_0^{-3}(\bar x = 0),
    \\ \Rightarrow f_0^{-3}(1)-f_0^{-3}(0)&=-\frac{\Delta p_c}{R_0}\lim\limits_{\bar x=0}\left(\frac{\partial}{\partial \bar x}\left(\frac{\bar x}{Q_t(-\Delta p_c\bar x)}\right)\right),\label{eq:f0_continuity}
\end{align}
which further restricts the choice of initial gap function $f_0$. 
Lastly, the target flow-rate $Q_t$ and critical pressure $\Delta p_c$ (Eq. \eqref{eq:Delta_p_c}) 
must be chosen in a way such that the gap height $f$ is real, positive and finite, i.e., $f^{-3}(\bar x)$ from Eq. \eqref{eq:f} should be positive and finite. 


\subsubsection{Examples of height profiles from target flow-rate pressure-drop characteristics}
\begin{figure}
    \centering
    \includegraphics[width=17.2cm]{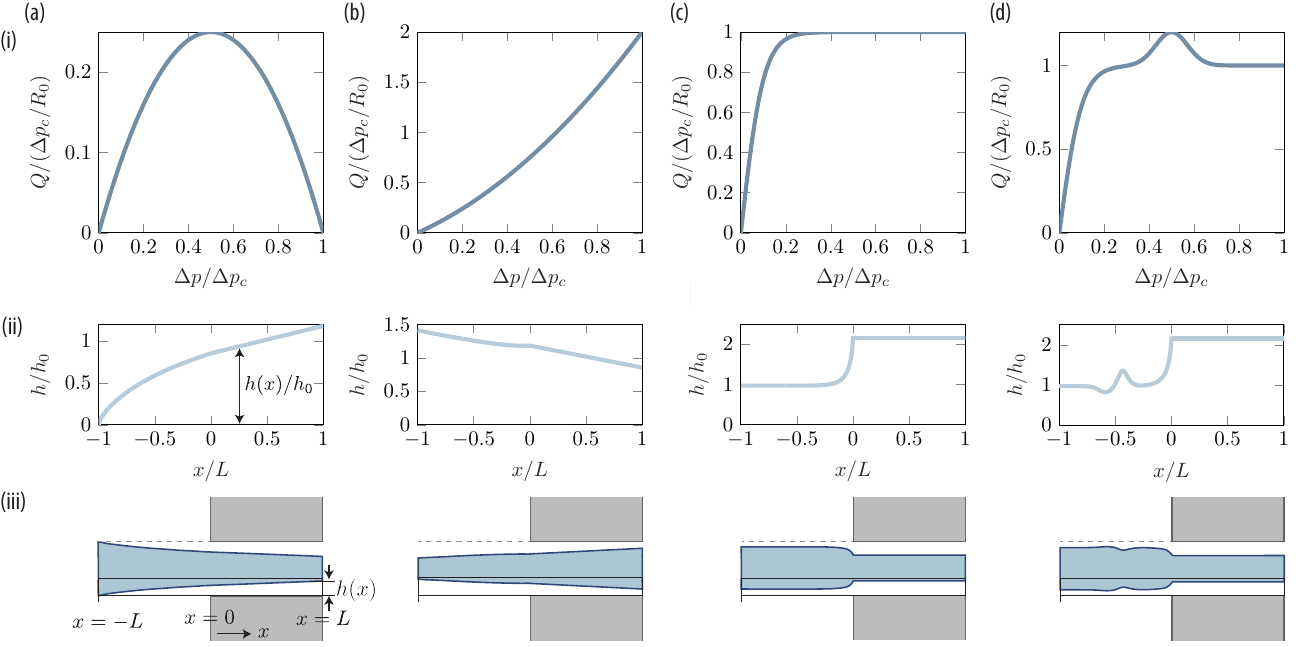}
    \caption{Examples of height profiles found for given flow-rate pressure-drop characteristics. (i) Target flow-rate pressure-drop characteristics. The flow rate is normalised by the flow through a pore with a constant height profile $h(x)=h_0$ at pressure difference $\Delta p=\Delta p_c$, i.e., by $\Delta p_c/R_0$ (Eq. \eqref{eq:R_0}), and the pressure is normalised by the characteristic pressure difference $\Delta p_c$ (Eq. \eqref{eq:Delta_p_c}). (ii) Corresponding normalised height profiles, $h(x)/h_0=f(x)$, from Eq. \eqref{eq:f} when the initial height profile is on the form of a linear profile (Eq. \eqref{eq:f0}). The axial coordiante $x$ is normalised by the pore length $L$. (iii) Schematics of the pore geometry. The box indicates the part of the geometry plotted in (ii) (not to scale). (a) Target flow rate profile from Eq. \eqref{eq:Qt_minus} with corresponding height profile from Eq. \eqref{eq:f_minus}. (b) Target flow rate profile from Eq. \eqref{eq:Qt_plus} with corresponding height profile from Eq. \eqref{eq:f_plus}. (c) Target flow rate profile from Eq. \eqref{eq:Qt_tanh} with corresponding height profile from Eq. \eqref{eq:f_tanh}. (d) Target flow rate profile from Eq. \eqref{eq:Qt_tanh_gauss} with corresponding height profile from Eq. \eqref{eq:f_tanh_gauss}.}
    \label{fig:fig3}
\end{figure}

In the following, we use one example of an initial height profile $f_0(\bar x)$ to show how height profiles for different target flow-rate pressure-drop characteristics can be determined. As we have two conditions that need to be satisfied for the final height profile to be continuous and to yield the desired flow-rate characteristic, we need the initial height function to have two free parameters that can be determined by the conditions in Eqs. \eqref{eq:f0_normalisation} and \eqref{eq:f0_continuity}. The initial height profile considered here is chosen as a linear function
\begin{align}
    f_0(\bar x)&=\alpha(1+\varepsilon \bar x),\label{eq:f0}
\end{align}
with parameters $\alpha$ and $\varepsilon$ determined from Eqs. \eqref{eq:f0_normalisation} and \eqref{eq:f0_continuity}. To illustrate the method, we consider the distinct examples shown in Fig. \ref{fig:fig3}(a-d). From left to right, they correspond to the target flow-rate pressure-drop relations
\begin{subequations}
\begin{align}
    Q_t(\Delta p)&=\frac{\Delta p}{R_0}\left(1-\frac{\Delta p}{\Delta p_c}\right),\label{eq:Qt_minus}
    \\ Q_t(\Delta p)&=\frac{\Delta p}{R_0}\left(1+\frac{\Delta p}{\Delta p_c}\right),\label{eq:Qt_plus}
    \\ Q_t(\Delta p)&=\frac{\Delta p_c}{R_0}\tanh\left(10\frac{\Delta p}{\Delta p_c}\right),\label{eq:Qt_tanh}\quad \text{and}
    \\ Q_t(\Delta p)&=\frac{\Delta p_c}{R_0}\left[\tanh\left(10\frac{\Delta p}{\Delta p_c}\right)+ \frac{1}{2\sqrt{2\pi}}\left(e^{-\frac{(\Delta p/\Delta p_c-1/2)^2}{(1/10)^2}}-e^{-\frac{(1/2)^2}{(1/10)^2}}\right)\right].\label{eq:Qt_tanh_gauss}
\end{align}
\end{subequations}
We note these cases increase linearly with $\Delta p$ around $\Delta p =0$ and are zero at zero pressure difference $Q_t(\Delta p=0)=0$. Eqs. \eqref{eq:Qt_minus} and \eqref{eq:Qt_plus} are worth highlighting, since they result in simple analytical solutions for the height profile, showing examples of valves that close and open, respectively. Eqs. \eqref{eq:Qt_tanh} and \eqref{eq:Qt_tanh_gauss} showcases systems that could be physiologically relevant, with regions of constant flow as a function of applied pressure drop, allowing for a constant supply of, e.g., nutrients. 

Inserting the flow-rate profiles from Eqs. \eqref{eq:Qt_minus}, \eqref{eq:Qt_plus}, \eqref{eq:Qt_tanh} and \eqref{eq:Qt_tanh_gauss} into Eqs. \eqref{eq:f}, \eqref{eq:f0_normalisation} and \eqref{eq:f0_continuity} we find their respective height profiles (Figs. \ref{fig:fig3} (a-d) (ii)),
\begin{subequations}
    \begin{align}
        f(\bar x) &=\left[\frac{1}{\alpha^3(1+\varepsilon(1+\bar x))^3}+\frac{1}{(1+\bar x)^2}\right]^{-1/3},\label{eq:f_minus}
        \\&\text{with}\quad \alpha=\frac{(2+\varepsilon)^{1/3}}{2^{1/3}(1+\varepsilon)^{2/3}}\quad\text{and}\quad \varepsilon = \frac{(118+3\sqrt{1509})^{2/3}-5(118+3\sqrt{1509})^{1/3}+7}{6(118+3\sqrt{1509})^{1/3}},\nonumber
        \\f(\bar x) &=\left[\frac{1}{\alpha^3(1+\varepsilon(1+\bar x))^3}-\frac{1}{(1-\bar x)^2}\right]^{-1/3},\label{eq:f_plus}
        \\&\text{with}\quad \alpha=\frac{(2+\varepsilon)^{1/3}}{2^{1/3}(1+\varepsilon)^{2/3}}\quad\text{and}\quad \varepsilon = \frac{(116+3\sqrt{1509})^{2/3}-7(116+3\sqrt{1509})^{1/3}-5}{6(116+3\sqrt{1509})^{1/3}},\nonumber
        \\f(\bar x)&=\left[\frac{1}{10}-\frac{1}{\tanh(10\bar x)}+\frac{10\bar x(1-\tanh^2(10\bar x))}{\tanh^2(10\bar x)}\right]^{-1/3},\label{eq:f_tanh}
        \\&\text{with}\quad \alpha=\alpha=\frac{5^{1/3}(2+\varepsilon)^{1/3}}{(1+\varepsilon)^{2/3}}\quad\text{and}\quad \varepsilon = 0,\nonumber
        \\f(\bar x)&=\left[\frac{1}{10}-\frac{1}{\tanh(10\bar x)-\frac{1}{2\sqrt{2\pi}}\left(e^{-\frac{(-\bar x-1/2)^2}{(1/10)^2}}-e^{-\frac{(1/2)^2}{(1/10)^2}}\right)}+\frac{10\bar x\left(1-\tanh^2(10\bar x)+\frac{20\bar x +10}{2\sqrt{2\pi}}e^{-\frac{(-\bar x-1/2)^2}{(1/10)^2}}\right)}{\left(-\tanh(10\bar x)+\frac{1}{2\sqrt{2\pi}}\left(e^{-\frac{(-\bar x-1/2)^2}{(1/10)^2}}-e^{-\frac{(1/2)^2}{(1/10)^2}}\right)\right)^2}\right]^{-1/3},\label{eq:f_tanh_gauss}
        \\&\text{with}\quad \alpha=\frac{5^{1/3}(2+\varepsilon)^{1/3}}{(1+\varepsilon)^{2/3}}\quad\text{and}\quad \varepsilon = 0.\nonumber
    \end{align}
\end{subequations}
Note that the values of $\alpha$ and $\varepsilon$ also refer to the parameters in the initial linear height profile (Eq. \eqref{eq:f0}). 

%
%

\subsubsection{Sensitivity analysis}
We end this section by discussing how sensitive the target flow-rate is to small variations in the derived height profile. If the system is to be used in an experiment, the fabrication process could result in small random variations in the height profile. 

\begin{figure}
    \centering
    \includegraphics[width=17.9cm]{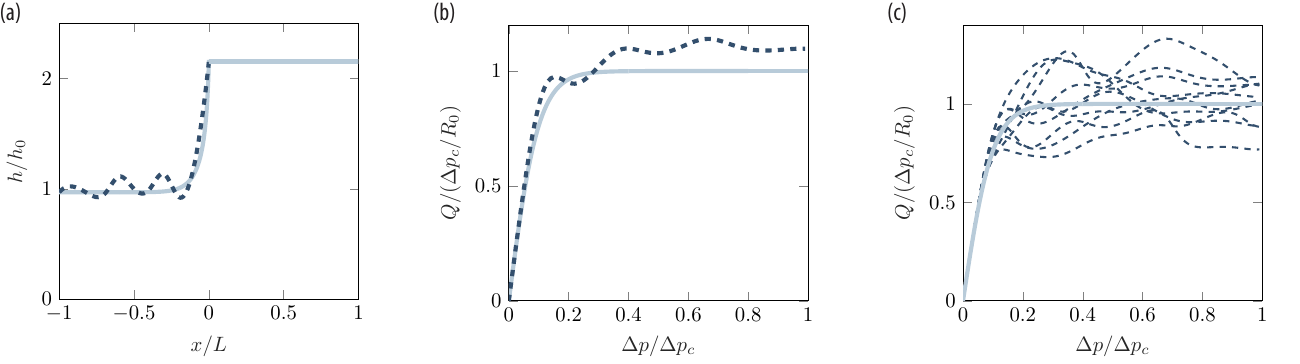}
    \caption{Sensitivity analysis. (a) Normalised height profile $h(x)/h_0$ from Eq. \eqref{eq:f_tanh} (light blue line) and with added random noise (dashed dark blue line). (b) Flow rate $Q$ as a function of pressure difference $\Delta p$ for the height profile with random noise from (a) (dashed dark blue line) and for the height profile without noise (solid light blue, Eq. \eqref{eq:Qt_tanh}). (c) Flow rate $Q$ as a function of pressure difference $\Delta p$ for ten iterations of random noise added to the height profile from Eq. \eqref{eq:f_tanh} (dashed dark blue lines) and for the height profile with no added noise (solid light blue line, Eq. \eqref{eq:Qt_tanh}). In both (b) and (c) the flow rate is normalised by the flow through a pore with a constant height profile $h(x)=h_0$ at pressure difference $\Delta p=\Delta p_c$, i.e., by $\Delta p_c/R_0$ (Eq. \eqref{eq:R_0}), and the pressure is normalised by the characteristic pressure difference $\Delta p_c$ (Eq. \eqref{eq:Delta_p_c}).}
    \label{fig:fig4}
\end{figure}

As an example, we consider the effect of adding random noise to the height profile from Eq. \eqref{eq:f_tanh}, which is derived from the hyperbolic tangent flow-rate profile (Eq. \eqref{eq:Qt_tanh}, Fig. \ref{fig:fig3} (c)). We add the noise by choosing nine equally spaced points between $x/L=-1$ and $x/L=0$ (not including $x/L=-1$ and $x/L=0$) and adding a random number to the height profile drawn from the standard normal distribution with a relative amplitude of $0.1$. The points are then connected using a spline function. An example of a height function with added random noise is seen in Fig. \ref{fig:fig4} (a). The corresponding flow-rate pressure-drop curve (dashed dark blue line in Fig. \ref{fig:fig4} (b)) deviates slightly from the desired target flow-rate (Eq. \eqref{eq:Qt_tanh} and solid light blue line in Fig. \ref{fig:fig4} (b)). By adding noise to the height profile for the target hyperbolic tangent flow-rate profile (Eq. \eqref{eq:f_tanh}) ten times, we see that the deviation from the target profile (Eq. \eqref{eq:Qt_tanh}) is approximately up to $25\%$ (Fig. \ref{fig:fig4} (c)).
%

\section{Discussion and conclusion}
In this paper we have presented a theoretical study of a sliding valve capable of generating a range of different flow-rate pressure-drop characteristics, depending on its geometry. The model system consists of a concentric cylindrical geometry, in which the radius of the inner cylinder can vary in the axial direction (Fig. \ref{fig:diagram}). When an applied pressure drop is present across the pore, the inner cylinder is displaced with respect to the outer cylinder, thereby changing the geometry of the pore. 

The flow-rate pressure-drop characteristics are found for two cases. First, for known height profiles of the gap between the two cylinders, the flow-rate as a function of applied pressure difference can be derived (Eq. \eqref{eq:Q_h}). A variety of behaviours can be observed depending on the gap height profile, where the system can both act as a check valve, only permitting flow below a certain critical pressure-drop, as well as increase super-linearly with applied pressure (Fig. \ref{fig:fig2}). 
And second, for known target flow-rate pressure-drop characteristics, the geometry of the pore can also be found by solving the inverse problem (Eq. \eqref{eq:f}), even for strongly non-linear fluid flow characteristics (Fig. \ref{fig:fig3}). 

Fluid flows interacting with movable and elastic elements are present in a variety of technical and biological systems. Controlling the fluid flow characteristics is important in, for example, lab-on-a-chip systems or biomedical applications \cite{oh2006review}, and the sliding valve system considered in this paper could potentially find applications here. Additionally, the sliding valve system may resemble that of plasmodesmata nanopores in plants, and could thus find applications in describing their pressure-driven fluid flow dynamics. 

\bibliography{refs.bib}

\begin{thebibliography}{16}%
\makeatletter
\providecommand \@ifxundefined [1]{%
 \@ifx{#1\undefined}
}%
\providecommand \@ifnum [1]{%
 \ifnum #1\expandafter \@firstoftwo
 \else \expandafter \@secondoftwo
 \fi
}%
\providecommand \@ifx [1]{%
 \ifx #1\expandafter \@firstoftwo
 \else \expandafter \@secondoftwo
 \fi
}%
\providecommand \natexlab [1]{#1}%
\providecommand \enquote  [1]{``#1''}%
\providecommand \bibnamefont  [1]{#1}%
\providecommand \bibfnamefont [1]{#1}%
\providecommand \citenamefont [1]{#1}%
\providecommand \href@noop [0]{\@secondoftwo}%
\providecommand \href [0]{\begingroup \@sanitize@url \@href}%
\providecommand \@href[1]{\@@startlink{#1}\@@href}%
\providecommand \@@href[1]{\endgroup#1\@@endlink}%
\providecommand \@sanitize@url [0]{\catcode `\\12\catcode `\$12\catcode
  `\&12\catcode `\#12\catcode `\^12\catcode `\_12\catcode `\%12\relax}%
\providecommand \@@startlink[1]{}%
\providecommand \@@endlink[0]{}%
\providecommand \url  [0]{\begingroup\@sanitize@url \@url }%
\providecommand \@url [1]{\endgroup\@href {#1}{\urlprefix }}%
\providecommand \urlprefix  [0]{URL }%
\providecommand \Eprint [0]{\href }%
\providecommand \doibase [0]{https://doi.org/}%
\providecommand \selectlanguage [0]{\@gobble}%
\providecommand \bibinfo  [0]{\@secondoftwo}%
\providecommand \bibfield  [0]{\@secondoftwo}%
\providecommand \translation [1]{[#1]}%
\providecommand \BibitemOpen [0]{}%
\providecommand \bibitemStop [0]{}%
\providecommand \bibitemNoStop [0]{.\EOS\space}%
\providecommand \EOS [0]{\spacefactor3000\relax}%
\providecommand \BibitemShut  [1]{\csname bibitem#1\endcsname}%
\let\auto@bib@innerbib\@empty
\bibitem [{\citenamefont {Choat}\ \emph {et~al.}(2008)\citenamefont {Choat},
  \citenamefont {Cobb},\ and\ \citenamefont {Jansen}}]{choat2008structure}%
  \BibitemOpen
  \bibfield  {author} {\bibinfo {author} {\bibfnamefont {B.}~\bibnamefont
  {Choat}}, \bibinfo {author} {\bibfnamefont {A.~R.}\ \bibnamefont {Cobb}},\
  and\ \bibinfo {author} {\bibfnamefont {S.}~\bibnamefont {Jansen}},\
  }\bibfield  {title} {\bibinfo {title} {Structure and function of bordered
  pits: new discoveries and impacts on whole-plant hydraulic function},\
  }\href@noop {} {\bibfield  {journal} {\bibinfo  {journal} {New Phytologist}\
  }\textbf {\bibinfo {volume} {177}},\ \bibinfo {pages} {608} (\bibinfo {year}
  {2008})}\BibitemShut {NoStop}%
\bibitem [{\citenamefont {Sotiropoulos}\ \emph {et~al.}(2016)\citenamefont
  {Sotiropoulos}, \citenamefont {Le},\ and\ \citenamefont
  {Gilmanov}}]{sotiropoulos2016fluid}%
  \BibitemOpen
  \bibfield  {author} {\bibinfo {author} {\bibfnamefont {F.}~\bibnamefont
  {Sotiropoulos}}, \bibinfo {author} {\bibfnamefont {T.~B.}\ \bibnamefont
  {Le}},\ and\ \bibinfo {author} {\bibfnamefont {A.}~\bibnamefont {Gilmanov}},\
  }\bibfield  {title} {\bibinfo {title} {Fluid mechanics of heart valves and
  their replacements},\ }\href@noop {} {\bibfield  {journal} {\bibinfo
  {journal} {Annual Review of Fluid Mechanics}\ }\textbf {\bibinfo {volume}
  {48}},\ \bibinfo {pages} {259} (\bibinfo {year} {2016})}\BibitemShut
  {NoStop}%
\bibitem [{\citenamefont {Brett}\ \emph {et~al.}(2011)\citenamefont {Brett},
  \citenamefont {Zhao}, \citenamefont {Stoia},\ and\ \citenamefont
  {Eddington}}]{brett2011controlling}%
  \BibitemOpen
  \bibfield  {author} {\bibinfo {author} {\bibfnamefont {M.-E.}\ \bibnamefont
  {Brett}}, \bibinfo {author} {\bibfnamefont {S.}~\bibnamefont {Zhao}},
  \bibinfo {author} {\bibfnamefont {J.~L.}\ \bibnamefont {Stoia}},\ and\
  \bibinfo {author} {\bibfnamefont {D.~T.}\ \bibnamefont {Eddington}},\
  }\bibfield  {title} {\bibinfo {title} {Controlling flow in microfluidic
  channels with a manually actuated pin valve},\ }\href@noop {} {\bibfield
  {journal} {\bibinfo  {journal} {Biomedical Microdevices}\ }\textbf {\bibinfo
  {volume} {13}},\ \bibinfo {pages} {633} (\bibinfo {year} {2011})}\BibitemShut
  {NoStop}%
\bibitem [{\citenamefont {Mosadegh}\ \emph {et~al.}(2010)\citenamefont
  {Mosadegh}, \citenamefont {Kuo}, \citenamefont {Tung}, \citenamefont
  {Torisawa}, \citenamefont {Bersano-Begey}, \citenamefont {Tavana},\ and\
  \citenamefont {Takayama}}]{mosadegh2010integrated}%
  \BibitemOpen
  \bibfield  {author} {\bibinfo {author} {\bibfnamefont {B.}~\bibnamefont
  {Mosadegh}}, \bibinfo {author} {\bibfnamefont {C.-H.}\ \bibnamefont {Kuo}},
  \bibinfo {author} {\bibfnamefont {Y.-C.}\ \bibnamefont {Tung}}, \bibinfo
  {author} {\bibfnamefont {Y.-s.}\ \bibnamefont {Torisawa}}, \bibinfo {author}
  {\bibfnamefont {T.}~\bibnamefont {Bersano-Begey}}, \bibinfo {author}
  {\bibfnamefont {H.}~\bibnamefont {Tavana}},\ and\ \bibinfo {author}
  {\bibfnamefont {S.}~\bibnamefont {Takayama}},\ }\bibfield  {title} {\bibinfo
  {title} {Integrated elastomeric components for autonomous regulation of
  sequential and oscillatory flow switching in microfluidic devices},\
  }\href@noop {} {\bibfield  {journal} {\bibinfo  {journal} {Nature Physics}\
  }\textbf {\bibinfo {volume} {6}},\ \bibinfo {pages} {433} (\bibinfo {year}
  {2010})}\BibitemShut {NoStop}%
\bibitem [{\citenamefont {Oh}\ and\ \citenamefont {Ahn}(2006)}]{oh2006review}%
  \BibitemOpen
  \bibfield  {author} {\bibinfo {author} {\bibfnamefont {K.~W.}\ \bibnamefont
  {Oh}}\ and\ \bibinfo {author} {\bibfnamefont {C.~H.}\ \bibnamefont {Ahn}},\
  }\bibfield  {title} {\bibinfo {title} {A review of microvalves},\ }\href@noop
  {} {\bibfield  {journal} {\bibinfo  {journal} {Journal of Micromechanics and
  Microengineering}\ }\textbf {\bibinfo {volume} {16}},\ \bibinfo {pages} {R13}
  (\bibinfo {year} {2006})}\BibitemShut {NoStop}%
\bibitem [{\citenamefont {Brandenbourger}\ \emph {et~al.}(2020)\citenamefont
  {Brandenbourger}, \citenamefont {Dangremont}, \citenamefont {Sprik},\ and\
  \citenamefont {Coulais}}]{brandenbourger2020tunable}%
  \BibitemOpen
  \bibfield  {author} {\bibinfo {author} {\bibfnamefont {M.}~\bibnamefont
  {Brandenbourger}}, \bibinfo {author} {\bibfnamefont {A.}~\bibnamefont
  {Dangremont}}, \bibinfo {author} {\bibfnamefont {R.}~\bibnamefont {Sprik}},\
  and\ \bibinfo {author} {\bibfnamefont {C.}~\bibnamefont {Coulais}},\
  }\bibfield  {title} {\bibinfo {title} {Tunable flow asymmetry and flow
  rectification with bio-inspired soft leaflets},\ }\href@noop {} {\bibfield
  {journal} {\bibinfo  {journal} {Physical Review Fluids}\ }\textbf {\bibinfo
  {volume} {5}},\ \bibinfo {pages} {084102} (\bibinfo {year}
  {2020})}\BibitemShut {NoStop}%
\bibitem [{\citenamefont {Skotheim}\ and\ \citenamefont
  {Mahadevan}(2004)}]{skotheim2004soft}%
  \BibitemOpen
  \bibfield  {author} {\bibinfo {author} {\bibfnamefont {J.}~\bibnamefont
  {Skotheim}}\ and\ \bibinfo {author} {\bibfnamefont {L.}~\bibnamefont
  {Mahadevan}},\ }\bibfield  {title} {\bibinfo {title} {Soft lubrication},\
  }\href@noop {} {\bibfield  {journal} {\bibinfo  {journal} {Physical review
  letters}\ }\textbf {\bibinfo {volume} {92}},\ \bibinfo {pages} {245509}
  (\bibinfo {year} {2004})}\BibitemShut {NoStop}%
\bibitem [{\citenamefont {Alvarado}\ \emph {et~al.}(2017)\citenamefont
  {Alvarado}, \citenamefont {Comtet}, \citenamefont {De~Langre},\ and\
  \citenamefont {Hosoi}}]{alvarado2017nonlinear}%
  \BibitemOpen
  \bibfield  {author} {\bibinfo {author} {\bibfnamefont {J.}~\bibnamefont
  {Alvarado}}, \bibinfo {author} {\bibfnamefont {J.}~\bibnamefont {Comtet}},
  \bibinfo {author} {\bibfnamefont {E.}~\bibnamefont {De~Langre}},\ and\
  \bibinfo {author} {\bibfnamefont {A.}~\bibnamefont {Hosoi}},\ }\bibfield
  {title} {\bibinfo {title} {Nonlinear flow response of soft hair beds},\
  }\href@noop {} {\bibfield  {journal} {\bibinfo  {journal} {Nature Physics}\
  }\textbf {\bibinfo {volume} {13}},\ \bibinfo {pages} {1014} (\bibinfo {year}
  {2017})}\BibitemShut {NoStop}%
\bibitem [{\citenamefont {Park}\ \emph {et~al.}(2018)\citenamefont {Park},
  \citenamefont {Tixier}, \citenamefont {Christensen}, \citenamefont
  {Arnbjerg-Nielsen}, \citenamefont {Zwieniecki},\ and\ \citenamefont
  {Jensen}}]{park2018viscous}%
  \BibitemOpen
  \bibfield  {author} {\bibinfo {author} {\bibfnamefont {K.}~\bibnamefont
  {Park}}, \bibinfo {author} {\bibfnamefont {A.}~\bibnamefont {Tixier}},
  \bibinfo {author} {\bibfnamefont {A.}~\bibnamefont {Christensen}}, \bibinfo
  {author} {\bibfnamefont {S.}~\bibnamefont {Arnbjerg-Nielsen}}, \bibinfo
  {author} {\bibfnamefont {M.}~\bibnamefont {Zwieniecki}},\ and\ \bibinfo
  {author} {\bibfnamefont {K.}~\bibnamefont {Jensen}},\ }\bibfield  {title}
  {\bibinfo {title} {Viscous flow in a soft valve},\ }\href@noop {} {\bibfield
  {journal} {\bibinfo  {journal} {Journal of Fluid Mechanics}\ }\textbf
  {\bibinfo {volume} {836}},\ \bibinfo {pages} {R3} (\bibinfo {year}
  {2018})}\BibitemShut {NoStop}%
\bibitem [{\citenamefont {Christov}\ \emph {et~al.}(2018)\citenamefont
  {Christov}, \citenamefont {Cognet}, \citenamefont {Shidhore},\ and\
  \citenamefont {Stone}}]{christov2018flow}%
  \BibitemOpen
  \bibfield  {author} {\bibinfo {author} {\bibfnamefont {I.~C.}\ \bibnamefont
  {Christov}}, \bibinfo {author} {\bibfnamefont {V.}~\bibnamefont {Cognet}},
  \bibinfo {author} {\bibfnamefont {T.~C.}\ \bibnamefont {Shidhore}},\ and\
  \bibinfo {author} {\bibfnamefont {H.~A.}\ \bibnamefont {Stone}},\ }\bibfield
  {title} {\bibinfo {title} {Flow rate--pressure drop relation for deformable
  shallow microfluidic channels},\ }\href@noop {} {\bibfield  {journal}
  {\bibinfo  {journal} {Journal of Fluid Mechanics}\ }\textbf {\bibinfo
  {volume} {841}},\ \bibinfo {pages} {267} (\bibinfo {year}
  {2018})}\BibitemShut {NoStop}%
\bibitem [{\citenamefont {Christensen}\ and\ \citenamefont
  {Jensen}(2020)}]{christensen2020viscous}%
  \BibitemOpen
  \bibfield  {author} {\bibinfo {author} {\bibfnamefont {A.~H.}\ \bibnamefont
  {Christensen}}\ and\ \bibinfo {author} {\bibfnamefont {K.~H.}\ \bibnamefont
  {Jensen}},\ }\bibfield  {title} {\bibinfo {title} {Viscous flow in a slit
  between two elastic plates},\ }\href@noop {} {\bibfield  {journal} {\bibinfo
  {journal} {Physical Review Fluids}\ }\textbf {\bibinfo {volume} {5}},\
  \bibinfo {pages} {044101} (\bibinfo {year} {2020})}\BibitemShut {NoStop}%
\bibitem [{\citenamefont {Oparka}\ and\ \citenamefont
  {Prior}(1992)}]{oparka1992direct}%
  \BibitemOpen
  \bibfield  {author} {\bibinfo {author} {\bibfnamefont {K.}~\bibnamefont
  {Oparka}}\ and\ \bibinfo {author} {\bibfnamefont {D.}~\bibnamefont {Prior}},\
  }\bibfield  {title} {\bibinfo {title} {Direct evidence for pressure-generated
  closure of plasmodesmata},\ }\href@noop {} {\bibfield  {journal} {\bibinfo
  {journal} {The Plant Journal}\ }\textbf {\bibinfo {volume} {2}},\ \bibinfo
  {pages} {741} (\bibinfo {year} {1992})}\BibitemShut {NoStop}%
\bibitem [{\citenamefont {Ruan}\ \emph {et~al.}(2001)\citenamefont {Ruan},
  \citenamefont {Llewellyn},\ and\ \citenamefont {Furbank}}]{ruan2001control}%
  \BibitemOpen
  \bibfield  {author} {\bibinfo {author} {\bibfnamefont {Y.-L.}\ \bibnamefont
  {Ruan}}, \bibinfo {author} {\bibfnamefont {D.~J.}\ \bibnamefont
  {Llewellyn}},\ and\ \bibinfo {author} {\bibfnamefont {R.~T.}\ \bibnamefont
  {Furbank}},\ }\bibfield  {title} {\bibinfo {title} {The control of
  single-celled cotton fiber elongation by developmentally reversible gating of
  plasmodesmata and coordinated expression of sucrose and k+ transporters and
  expansin},\ }\href@noop {} {\bibfield  {journal} {\bibinfo  {journal} {The
  Plant Cell}\ }\textbf {\bibinfo {volume} {13}},\ \bibinfo {pages} {47}
  (\bibinfo {year} {2001})}\BibitemShut {NoStop}%
\bibitem [{\citenamefont {Nicolas}\ \emph {et~al.}(2017)\citenamefont
  {Nicolas}, \citenamefont {Grison}, \citenamefont {Tr{\'e}pout}, \citenamefont
  {Gaston}, \citenamefont {Fouch{\'e}}, \citenamefont {Cordeli{\`e}res},
  \citenamefont {Oparka}, \citenamefont {Tilsner}, \citenamefont {Brocard},\
  and\ \citenamefont {Bayer}}]{nicolas2017architecture}%
  \BibitemOpen
  \bibfield  {author} {\bibinfo {author} {\bibfnamefont {W.~J.}\ \bibnamefont
  {Nicolas}}, \bibinfo {author} {\bibfnamefont {M.~S.}\ \bibnamefont {Grison}},
  \bibinfo {author} {\bibfnamefont {S.}~\bibnamefont {Tr{\'e}pout}}, \bibinfo
  {author} {\bibfnamefont {A.}~\bibnamefont {Gaston}}, \bibinfo {author}
  {\bibfnamefont {M.}~\bibnamefont {Fouch{\'e}}}, \bibinfo {author}
  {\bibfnamefont {F.~P.}\ \bibnamefont {Cordeli{\`e}res}}, \bibinfo {author}
  {\bibfnamefont {K.}~\bibnamefont {Oparka}}, \bibinfo {author} {\bibfnamefont
  {J.}~\bibnamefont {Tilsner}}, \bibinfo {author} {\bibfnamefont
  {L.}~\bibnamefont {Brocard}},\ and\ \bibinfo {author} {\bibfnamefont {E.~M.}\
  \bibnamefont {Bayer}},\ }\bibfield  {title} {\bibinfo {title} {Architecture
  and permeability of post-cytokinesis plasmodesmata lacking cytoplasmic
  sleeves},\ }\href@noop {} {\bibfield  {journal} {\bibinfo  {journal} {Nature
  Plants}\ }\textbf {\bibinfo {volume} {3}},\ \bibinfo {pages} {1} (\bibinfo
  {year} {2017})}\BibitemShut {NoStop}%
\bibitem [{\citenamefont {Park}\ \emph {et~al.}(2019)\citenamefont {Park},
  \citenamefont {Knoblauch}, \citenamefont {Oparka},\ and\ \citenamefont
  {Jensen}}]{park2019controlling}%
  \BibitemOpen
  \bibfield  {author} {\bibinfo {author} {\bibfnamefont {K.}~\bibnamefont
  {Park}}, \bibinfo {author} {\bibfnamefont {J.}~\bibnamefont {Knoblauch}},
  \bibinfo {author} {\bibfnamefont {K.}~\bibnamefont {Oparka}},\ and\ \bibinfo
  {author} {\bibfnamefont {K.~H.}\ \bibnamefont {Jensen}},\ }\bibfield  {title}
  {\bibinfo {title} {Controlling intercellular flow through mechanosensitive
  plasmodesmata nanopores},\ }\href@noop {} {\bibfield  {journal} {\bibinfo
  {journal} {Nature Communications}\ }\textbf {\bibinfo {volume} {10}},\
  \bibinfo {pages} {3564} (\bibinfo {year} {2019})}\BibitemShut {NoStop}%
\bibitem [{\citenamefont {Yang}\ \emph {et~al.}(2022)\citenamefont {Yang},
  \citenamefont {Dong}, \citenamefont {Wang}, \citenamefont {Liu},
  \citenamefont {Li},\ and\ \citenamefont {Li}}]{yang2022wearable}%
  \BibitemOpen
  \bibfield  {author} {\bibinfo {author} {\bibfnamefont {Z.}~\bibnamefont
  {Yang}}, \bibinfo {author} {\bibfnamefont {L.}~\bibnamefont {Dong}}, \bibinfo
  {author} {\bibfnamefont {M.}~\bibnamefont {Wang}}, \bibinfo {author}
  {\bibfnamefont {G.}~\bibnamefont {Liu}}, \bibinfo {author} {\bibfnamefont
  {X.}~\bibnamefont {Li}},\ and\ \bibinfo {author} {\bibfnamefont
  {Y.}~\bibnamefont {Li}},\ }\bibfield  {title} {\bibinfo {title} {A wearable
  insulin delivery system based on a piezoelectric micropump},\ }\href@noop {}
  {\bibfield  {journal} {\bibinfo  {journal} {Sensors and Actuators A:
  Physical}\ }\textbf {\bibinfo {volume} {347}},\ \bibinfo {pages} {113909}
  (\bibinfo {year} {2022})}\BibitemShut {NoStop}%
\end{thebibliography}%
\end{document}